\documentclass[%
 reprint,
superscriptaddress,
 amsmath,amssymb,
 aps,
]{revtex4-1}

\usepackage{graphicx}
\usepackage{dcolumn}
\usepackage{multirow}
\usepackage{bm}
\usepackage[colorlinks= true, linkcolor=blue, citecolor=blue, urlcolor=blue]{hyperref}
\usepackage{lipsum}
\usepackage{bbold}
\usepackage{mathtools}
\usepackage[normalem]{ulem}
\usepackage{hyperref}

\def\bea{\begin{eqnarray}}
\def\eea{\end{eqnarray}}

\usepackage{xcolor}

\newcommand{\eref}[1]{Eq.~(\ref{#1})}
\newcommand{\fref}[1]{Fig.~\ref{#1}} 
 
\begin{document}
\newcommand{\aref}[1]{Appendix ~\ref{#1}}%

\title{Current fluctuations in finite-sized one-dimensional\\ non-interacting passive and active systems}

\author{Arup Biswas}
\email{arupb@imsc.res.in}
\affiliation{The Institute of Mathematical Sciences, CIT Campus, Taramani, Chennai 600113, India \& Homi Bhabha National Institute, Training School Complex, Anushakti Nagar, Mumbai 400094, India}
\author{Stephy Jose}
\email{stephyjose@tifrh.res.in}
\affiliation{Tata Institute of Fundamental Research, Hyderabad 500046, India}
\author{Arnab Pal}
\email{arnabpal@imsc.res.in}
\affiliation{The Institute of Mathematical Sciences, CIT Campus, Taramani, Chennai 600113, India \& Homi Bhabha National Institute, Training School Complex, Anushakti Nagar, Mumbai 400094, India}
\author{Kabir Ramola}
\email{kramola@tifrh.res.in}
\affiliation{Tata Institute of Fundamental Research, Hyderabad 500046, India}

\begin{abstract}
We investigate the problem of effusion of particles initially confined in a finite one-dimensional box of size $L$. We study both passive as well active scenarios, involving non-interacting diffusive particles and run-and-tumble particles, respectively. We derive analytic results for the fluctuations in the number of particles exiting the boundaries of the finite confining box. The statistical properties of this quantity crucially depend on how the system is prepared initially. Two common types of averages employed to understand the impact of initial conditions in stochastic systems are annealed and quenched averages. It is well known that for an infinitely extended system, these different initial conditions produce quantitatively different fluctuations, even in the infinite time limit. We demonstrate explicitly that in finite systems, annealed and quenched fluctuations become equal beyond a system-size dependent timescale, $t \sim L^2$. For diffusing particles, the fluctuations exhibit a $\sqrt{t}$ growth at short times and decay as $1/\sqrt{t}$ for time scales, $t \gg L^2/D$, where $D$ is the diffusion constant. Meanwhile, for run-and-tumble particles, the fluctuations grow linearly at short times and then decay as $1/\sqrt{t}$ for time scales, $t \gg L^2/D_{\text{eff}}$, where $D_{\text{eff}}$ represents the effective diffusive constant for run-and-tumble particles. To study the effect of confinement in detail, we also analyze two different setups (i) with one reflecting boundary and (ii) with both boundaries open. 

\end{abstract}

\pacs{Valid PACS appear here}
\maketitle

\section{Introduction} The study of the effect of initial conditions on the transport properties of stochastic systems has attracted considerable interest in the past years~\cite{le1989annealed,van1991fluctuations,derrida2004current,derrida2009current2,ferrari2010interacting,lizana2010foundation,krapivsky2012fluctuations,derridaprl2004,leibovich2013everlasting,krapivsky2014large,banerjee2022role,mallick2022exact,dandekar2022dynamical,dean2023effusion}. Notably, these studies have revealed that the distributions of quantities such as the tracer particle displacement or the integrated current across a region are different depending on the initial condition involving the positions of particles~\cite{mallick2022exact,dandekar2022dynamical,dean2023effusion,di2023current}.
Two ensembles of initial conditions that are commonly used to study this effect are (i) annealed setting, which allows for random fluctuations in the initial condition, and (ii) quenched setting, where the initial condition is deterministic~\cite{derrida2009current,cividini2017tagged,banerjee2022role,krapivsky2014large}. To gain an initial understanding of the relevance of initial conditions, imagine a set of particles initially confined in a one-dimensional channel, free to diffuse. Several intriguing questions arise: Does a static disorder in the initial arrangement of particles influence the dynamic behavior of the system? Furthermore, does this effect persist over large times, particularly when the channel length is finite? What happens if there is an asymmetry in the boundary conditions of this confining channel?

Previous studies have extensively examined the problem of effusion using model systems that are infinitely extended. One such model considers a semi-infinite confining channel bounded between $x \in (-\infty,0]$, where the fluctuations in the number of particles crossing the origin $x=0$ up to time $t$ are investigated~\cite{banerjee2020current,dean2023effusion}. For the case of diffusive particles, the annealed setting exhibits larger fluctuations by a factor of $\sqrt{2}$ as compared to the quenched setting~\cite{leibovich2013everlasting,derrida2009current2,krapivsky2012fluctuations,dean2023effusion,banerjee2020current}. Even for non-interacting active particles, the annealed setting exhibits larger fluctuations by a factor of $\sqrt{2}$ at large times, as the dynamics effectively becomes diffusive~\cite{banerjee2020current,jose2023generalized,jose2023effect}. 

\begin{figure*}
    \centering
    \includegraphics[width=17.5cm]{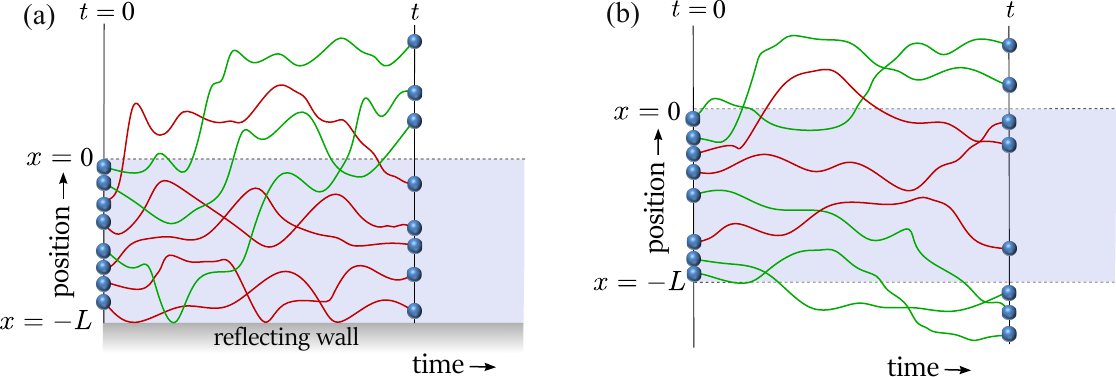}
    \caption{Schematic representation of the trajectories of $N=8$ non-interacting particles, initially confined in a box bounded between $[-L,0]$. We are interested in the number of particles present outside this box at time $t$, denoted as $Q(t)$. The green trajectories indicate those contributing to a non-zero $Q(t)$, while the red ones do not contribute. We examine two setups:~(a)~In the left panel, a reflecting wall is present at $x=-L$, allowing particles to escape only through the origin $x=0$.~(b)~In the right panel, both sides are open, enabling particles to exit through either boundary. In the illustrated figure, $Q(t)=3$ for the former case and $Q(t)=5$ for the latter. 
    }
    \label{fig_schematic}
\end{figure*}

In this paper, we focus on the dynamic properties of the particle flux $Q$ across the boundaries of a \textit{finite} confining box bounded between $x \in [-L,0]$. We specifically investigate the interplay between the initial conditions and the system geometry on the fluctuations of $Q$. We consider two setups: (i) with a reflecting boundary condition at $-L$, and (ii) with both the boundaries at $-L$ and $0$ open. Interestingly, we demonstrate that in both these cases, annealed and quenched fluctuations converge and become equal at a timescale determined by the system size $L$ and the parameters of the model studied. For diffusing particles, the fluctuations exhibit a $\sqrt{t}$ growth at short times and decay as $1/\sqrt{t}$ for time scales $t \gg L^2/D$, where $D$ represents the diffusion constant. Meanwhile, for run-and-tumble particles~\cite{malakar2018steady,evans2018run,mori2020universal,mori2020universalp,singh2019generalised,angelani2014first,martens2012probability,mori2021condensation,jose2022active,jose2022first,mallikarjun2023chiral}, the fluctuations grow linearly at short times and then decay as $1/\sqrt{t}$ for time scales $t \gg L^2/D_{\text{eff}}$, where $D_{\text{eff}}$ denotes the effective diffusive constant for run-and-tumble particles. For diffusive systems, the ratio of the fluctuations due to annealed and quenched initial conditions changes from a value of $\sqrt{2}$ (which is equal to the ratio observed in an infinite system) at short times to $1$ at large times for both the geometries; for active particles, it changes from the value of $2$ (infinite system) at short times to $1$ at large times in a similar vein. Intriguingly, we also show that the boundary conditions of the confining box play a crucial role in determining the dynamic behavior of $Q$. The setup with two open boundaries displays larger fluctuations by a factor of $2$ at short times and smaller fluctuations by a factor of $1/2$ at large times as compared to the setup with one open boundary for both passive and active cases.

The paper is organized as follows. In Sec.~\ref{formalism}, we introduce the models that we use to study the fluctuations in the particle flux $Q$. In Sec.~\ref{sec_diffusion} and Sec.~\ref{sec_rtp}, we present exact analytical results for the fluctuations in both diffusive and active systems. We present the conclusions from the study in Sec.~\ref{sec_discussion}. Finally, we present details related to some of the calculations in Appendices~\ref{appendix_diffusion} and~\ref{appendix_rtp}.


\section{The formalism}
\label{formalism}


In this section, we generalize the formalism developed in \cite{banerjee2020current} to study the fluctuations in the current of particles across the boundaries of a finite-sized box. We consider $N$ non-interacting particles initially distributed with a uniform density $\rho=N/L$ in a finite one-dimensional box bounded between $[-L,0]$. These particles evolve over time following their underlying dynamics such as diffusion or run-and-tumble motion. The quantity of interest is the number of particles exiting the boundaries of the box up to time $t$, equivalent to the number of particles present outside the box at time $t$ (see \fref{fig_schematic} for a schematic representation). We denote this quantity as $Q(t)$, representing the flux or integrated current through the boundaries of the box up to time $t$. The number of particles present outside the box can be expressed using an indicator function $\mathcal{I}(t)$, defined as
\begin{equation}
    {\cal I}_i(t)= 
\begin{cases}
   1,&\text{if the $i{\text {th}}$ particle is outside $[-L,0]$ at $t$},\\
     0,&\text{otherwise} .
\end{cases}
\end{equation} 
The current $Q(t)$ is then given as
\begin{align}
    Q(t)=\sum_{i=1}^N \mathcal{I}_i(t).
\end{align}
We are primarily interested in the statistical properties of the random variable $Q$. Generally, two sources of randomness are associated with the measurement of $Q$; the randomness in the initial positions of the particles and the randomness due to the inherent stochasticity of the underlying dynamics of the particles. There are two distinct methods for averaging over these sources of randomness: (i) the annealed average - which corresponds to simultaneous averaging over all initial conditions and noise history (ii) the quenched average - where one first averages over noise history for a fixed initial realization, followed by averaging over all possible initial realizations. The formal definitions of these averages are provided in the subsequent sections.

Let us denote by $\{x_i\}$ a distinct set of initial positions of the particles. For the fixed initial positions $\{x_i\}$, the probability distribution of $Q$ is given as
\begin{align}
    P(Q,t,\{x_i\})=\left\langle \delta \left(Q-\sum_{i=1}^N \mathcal{I}_i(t)\right)\right\rangle_{\{x_i\}}.
\end{align}
The angular bracket $\langle...\rangle_{\{x_i\}}$ in the above expression denotes an average over all trajectories of the particles for a fixed initial condition $\{x_i\}$. Moving forward it will be convenient to work with the moment-generating function of $Q$ defined as 
\begin{align}
    \sum_{Q=0}^{\infty} e^{-pQ}P(Q,t,\{x_i\})&=\langle e^{-pQ}\rangle_{\{x_i\}}\nonumber\\
    &=\left\langle \text{exp} \left(-p\sum_{i=1}^N \mathcal{I}_i(t)\right)\right\rangle_{\{x_i\}}.
\end{align}
We next use the identity $e^{-p\mathcal{I}_i(t)}=1-(1-e^{-p})\mathcal{I}_i(t)$ and the independent nature of the dynamics of the particles to obtain
\begin{align}
    \langle e^{-pQ}\rangle_{\{x_i\}}= \prod_{i=1}^N\left[1-(1-e^{-p})\langle \mathcal{I}_i(t)\rangle_{\{x_i\}}\right]
    \label{gf}.
\end{align}
Here $\langle \mathcal{I}_i(t)\rangle_{\{x_i\}}$  represents the probability that the $i$th particle is present outside the region $x\in[-L,0]$ at time $t$.  Depending on the underlying dynamics and the geometry of the system under consideration, this quantity will be different. We study two different cases where (i) there is a reflecting boundary at $x=-L$ (see \fref{fig_schematic}~(a)) and (ii) when both the boundaries at $x=0$ and $x=-L$ are open~(see \fref{fig_schematic}~(b)). In the first case, particles exit only through the boundary at $x=0$, however, in the latter case, they exit either through $x=0$ or $x=-L$. Denoting the expectation $\langle \mathcal{I}_i(t)\rangle_{\{x_i\}}$ by $U(x_i,t)$, we obtain 
\begin{align}
   U(x_i,t)= \int_0^\infty G(x,t|x_i)dx, \label{u-ref}
\end{align}
for the case with a reflecting boundary
and 
\begin{align}
   U(x_i,t)= \int_{-\infty}^{-L} G(x,t|x_i)dx+\int_0^\infty G(x,t|x_i)dx, \label{u}
\end{align}
when both boundaries are open. Here, $G(x,t|x_i)$ is the Green's function defined as the probability density to find a particle at a position  $x$ at time $t$ starting from the position $x_i$ at time $t=0$. From Eq.~\eqref{gf}, we obtain the expression for the generating function of $Q$ as
\begin{align}
    \langle e^{-pQ}\rangle_{\{x_i\}}= \prod_{i=1}^N\left[1-(1-e^{-p})U(x_i,t)\right],\label{gf-u}
\end{align}
where the expressions for the function $U$ for the settings with reflecting boundary and open boundaries are given in Eqs.~\eqref{u-ref} and~\eqref{u} respectively. The average over the initial conditions $\{x_i\}$ can now be done in two ways, as discussed below.

\subsection{Annealed setting}
Let us denote by the symbol $\overline{(...)}$ as an average over the initial conditions on the positions of the particles. Performing an average over the initial positions in \eref{gf-u}, we obtain
\begin{align}
    \overline{\langle e^{-pQ}\rangle_{\{x_i\}}}= \prod_{i=1}^N\left[1-(1-e^{-p})\overline{U(x_i,t)}\right].
\end{align}
Since the position of each particle is distributed independently according to a uniform distribution in the interval $x_i\in [-L,0]$, this expectation can be further simplified to
\begin{align}
      \overline{\langle e^{-pQ}\rangle_{\{x_i\}}}&= \prod_{i=1}^N\left[1-(1-e^{-p})\frac{1}{L}\int_{-L}^0 U(x_i,t)dx_i\right]\nonumber \\
    &=\left[1-(1-e^{-p})\frac{1}{L}\int_{-L}^0 U(z,t)dz\right]^N,
    \label{gf_annealed}
\end{align}
where we have assigned a general variable $z\equiv x_i$ as the motion of the particles is independent. Defining $P_{\text{an}}(Q,t)$ as the probability distribution for $Q$ in the annealed setting, we have
\begin{align}
     \sum_{Q=0}^{\infty} e^{-pQ}P_{\text{\text{an}}}(Q,t)=\overline{\langle e^{-pQ}\rangle_{\{x_i\}}}.\label{pan}
\end{align} 

For finite $N,~L$, a small $p$ expansion of Eq.~\eqref{gf_annealed} yields the expressions for the first few moments from which we can obtain the expressions for the mean $\mu_{\text{an}}(L,t)$ and the variance $\sigma^2_{\text{an}}(L,t)$ of $Q$ as
\begin{align}
   \mu_{\text{an}}(L,t) &=\overline{\langle Q \rangle}=\langle Q \rangle_{\text{an}}~, \label{mean-an} \\
   \label{var_two_walls_an}
      \sigma^2_{\text{an}}(L,t)&=\overline{\langle Q^2 \rangle}-{\overline{\langle Q \rangle}^2} \nonumber\\
      &=\langle Q^2 \rangle_{\text{an}} - {\langle Q \rangle_{\text{an}}}^2\nonumber\\
      &=\mu_{\text{an}}(L,t)-\frac{1}{\rho L}\mu_{\text{an}}^2(L,t).
\end{align}
In the above expression, we have replaced $N$ by $\rho L$. The quantity $\mu_{\text{an}}(L,t)$ can be computed as
\begin{equation}
     \mu_{\text{an}}(L,t)=\rho \int_{-L}^0 U(z,t)dz,
    \label{mean_two_walls}
\end{equation}
 where the expression for $U(z,t)$ is given in \eref{u-ref} and \eref{u} for the cases with one and two open boundaries respectively. 
So far, the majority of studies on the dynamic behavior of $Q$ have focused on infinite systems ($L\to \infty$). For an infinitely extended system with non-interacting particles, the mean and the variance are the same in the annealed setting. However, as we observe from Eqs.~\eqref{mean-an}~and~\eqref{var_two_walls_an},~ they are not identical when the system size is finite and furthermore, there is a $L$ dependent correction term in the variance. In the limit $L\to \infty$ one recovers the known result, $\mu_{\text{an}} (L\to \infty,t)=\sigma^2_{\text{an}}(L\to \infty,t)$~\cite{banerjee2020current}. 

\subsection{Quenched setting}
In the quenched setting, we first perform an average over the trajectories for a fixed initial condition and then average over the initial conditions of the system with a mean density $\rho$. The generating function for $Q$ in the quenched setting can be mathematically computed as
\begin{align}
    \sum_{Q=0}^{\infty}P_{\text{qu}}(Q,t)e^{-pQ}=\exp[\overline{\text{ln}\langle e^{-pQ}\rangle_{\{x_i\}}}].
\end{align}
Taking a logarithm of both sides of \eref{gf-u}, we obtain
\begin{align}
    \text{ln}\langle e^{-pQ}\rangle_{\{x_i\}}=\sum_{i=1}^N\ln [1-(1-e^{-p})U(x_i,t)].
\end{align}
Next performing an average over the initial positions in the above equation yield
\begin{align}
    \overline{\text{ln}\langle e^{-pQ}\rangle_{\{x_i\}}}&=\sum_{i=1}^N \frac{1}{L} \int_{-L}^0\ln [1-(1-e^{-p})U(x_i,t)] dx_i \nonumber\\
    &=\frac{N}{L} \int_{-L}^0\ln [1-(1-e^{-p})U(z,t)] dz \nonumber\\
    &=\rho \int_{-L}^0\ln [1-(1-e^{-p})U(z,t)] dz \nonumber\\
    &=I(p,t),
\end{align}
where 
\begin{align}
    I(p,t)=\rho \int_{-L}^0\ln [1-(1-e^{-p})U(z,t)] dz .
    \label{ip}
\end{align}
Finally, we obtain the expression for the generating function for the distribution of $Q$ in the quenched setting as
\begin{align}
    \sum_{Q=0}^{\infty}P_{\text{qu}}(Q,t)e^{-pQ}=\exp[I(p,t)].\label{pqu}
\end{align}

Performing a small $p$ expansion and collecting the terms at first and second orders of $p$, we obtain the expression for the mean $\mu_{\text{qu}}(L,t)$ and the variance  $\sigma^2_{\text{qu}}(L,t)$ of $Q$ in the quenched setting as
\begin{align}
   \mu_{\text{qu}}(L,t)&=\overline{\langle Q \rangle}=\langle Q \rangle_{\text{qu}}=\mu_{\text{an}}(L,t) , \label{mean_two_walls_qu} \\
      \sigma^2_{\text{qu}}(L,t)&=\overline{\langle Q^2 \rangle-\langle Q \rangle^2}\nonumber\\
      &=\langle Q^2 \rangle_{\text{qu}} - {\langle Q \rangle_{\text{qu}}}^2\nonumber\\ 
     &=\mu_{\text{qu}}(L,t)-\rho \int_{-L}^0dz~ U^2(z,t),
     \label{var_two_walls_qu}
\end{align}
where $\mu_{\text{qu}}(L,t)=\mu_{\text{an}}(L,t)$ is given by \eref{mean_two_walls}. 
The mean in the annealed and quenched settings are the same even when the system size is finite. However, the higher-order cumulants are different. 

    \label{an-qu-schematic}

In what follows, we study two specific examples of a system of diffusive particles and active run-and-tumble particles.


\begin{table}[t!]
\centering
{\setlength{\extrarowheight}{7pt}
\begin{tabular}{c|ccc}
&$t \rightarrow 0$ & $t \rightarrow \infty$          \\ [7pt]
\hline 
 \multirow{2}{*}{Diffusion reflecting boundary} & $ \frac{ \rho  \sqrt{D t}}{\sqrt{\pi }}$   & $\frac{\rho L^2}{\sqrt{\pi }   \sqrt{D t}}$   & $\sigma^{\text{diff}}_{\text{an}}(L,t)^2$  \\ [7pt]
 \cline{2-4}
&  $\frac{1}{\sqrt{2}}\frac{ \rho  \sqrt{D t}}{\sqrt{ \pi }}$   & $\frac{\rho L^2}{ \sqrt{\pi }   \sqrt{D t}}$   & $\sigma^{\text{diff}}_{\text{qu}}(L,t)^2$  \\ [7pt] 
   \hline
  \multirow{2}{*}{Diffusion finite interval} & $2\frac{ \rho  \sqrt{D t}}{\sqrt{\pi }}$   & $\frac{1}{2}\frac{\rho L^2}{ \sqrt{\pi }   \sqrt{D t}}$   & $\sigma^{\text{diff}}_{\text{an}}(L,t)^2$  \\ [7pt] 
       \cline{2-4}
 & $\sqrt{2}\frac{ \rho  \sqrt{D t}}{\sqrt{\pi }}$  & $\frac{1}{2}\frac{\rho L^2}{ \sqrt{\pi }  \sqrt{D t}}$ & $\sigma^{\text{diff}}_{\text{qu}}(L,t)^2$   \\ 
\end{tabular}}
 \caption{Asymptotic behavior of current fluctuations for diffusive motion in the annealed and quenched settings.}
\label{table_BM}
\end{table}

\section{Diffusive particles}
\label{sec_diffusion}

In this section, we consider a set of diffusive particles initially confined in a finite one-dimensional box bounded between $[-L,0]$. We also consider two distinct set-ups; one in the presence of a reflecting boundary and the other with both boundaries open. We summarize the asymptotic behaviors of current fluctuations for both these cases in Table~\ref{table_BM}. 

We first focus on the case with a reflecting wall at $x=-L$.
\begin{figure*}
    \centering
    \includegraphics[width=17.5cm]{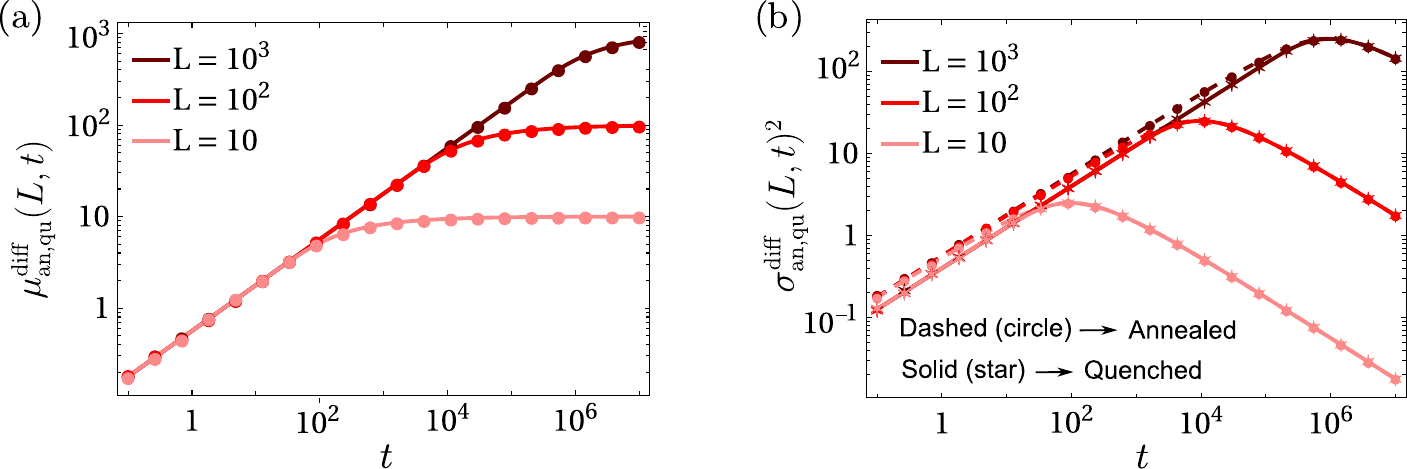}
    \caption{Behavior of the~(a)~mean and the~(b)~variance of current through the origin in the presence of a reflecting wall at $x=-L$ for $N$ diffusive Brownian particles. The mean in the annealed and quenched settings are the same. The variance in the quenched setting (solid curves) differs from the annealed (dashed curves) by a factor of $\sqrt{2}$ at times $t\ll L^2/D$, however, they become equal at times $t\gg L^2/D$. The parameter values used are $\rho=1, D=1$. The stars and circles represent the results obtained through numerical simulations of the microscopic model for the quenched and annealed settings respectively.}
    \label{fig:diff_ref}
\end{figure*}

\subsection{One reflecting wall}
In this section, we study the scenario where the boundary at $x=-L$ is a reflecting wall. The Green's function for a single diffusive Brownian particle in this case can be derived as~\cite{feller1991introduction}
\begin{align}
  G(x,t|x_i)=  \frac{1}{\sqrt{4 \pi D t}}\left(e^{-\frac{(2 L+x+x_i)^2}{4 D t}}+e^{-\frac{(x-x_i)^2}{4D t}}\right),
\end{align}
which can be substituted in \eref{u-ref} to obtain
\begin{align} \label{u-ref-diff}
   U(x_i,t)=  \frac{1}{2} \left(1+\text{erf}\left(\frac{x_i}{2 \sqrt{D t}}\right)+\text{erfc}\left(\frac{2 L+x_i}{2 \sqrt{D t}}\right)\right),
\end{align}
where \text{erf}($z$) and \text{erfc}($z$) are the error function and complementary error function, respectively. Having obtained the expression for $U(x_i,t)$, we can now compute the expressions for the mean and the variance of $Q$ for both annealed and quenched settings as detailed below.

\subsubsection{Annealed setting}
Substituting Eq.~\eqref{u-ref-diff} in Eq.~\eqref{mean_two_walls}, we obtain the expression for the mean of $Q$ for a system of diffusing particles in the annealed setting as
\begin{align}
    &\mu^{\text{diff}}_{\text{an}} (L,t)\nonumber\\
    &=\underbrace{\frac{\rho\sqrt{Dt}}{\sqrt{\pi}}}_{\text{infinite size limit}}+ \underbrace{\rho L\left(\text{erfc}\left(\frac{L}{ \sqrt{D t}}\right)-\frac{\sqrt{Dt}}{\sqrt{\pi}L}e^{-\frac{L^2}{Dt}}\right)}_{\text{finite size correction}} .\label{mean-diff-an-ref}
\end{align}
The first term in the mean does not have any explicit dependence on the system size $L$. This is the result one expects in the case of an infinite system size limit ($L\to\infty$). The second term in the parentheses contains the finite size corrections which vanishes in the limit $L\to\infty$. The expression for the variance follows from \eref{var_two_walls_an} as
\begin{align}
    {\sigma^{\text{diff}}_{\text{an}}(L,t)}^2&=\mu^{\text{diff}}_{\text{an}}(L,t)-\frac{1}{\rho L}{\mu^{\text{diff}}_{\text{an}}(L,t)}^2,
    \label{var-an-diff-ref}
\end{align}
with $\mu^{\text{diff}}_{\text{an}}(L,t)$ given by \eref{mean-diff-an-ref}. Since the exact expression for the variance is quite lengthy, we do not provide it here. \fref{fig:diff_ref} shows the behavior of the mean and the variance as a function of time obtained from Eqs.~\eqref{mean-diff-an-ref}~and~\eqref{var-an-diff-ref} for different system sizes keeping the density $\rho=1$ fixed.


Both the mean and variance increase monotonically with time for $t\ll L^2/D$. Taking this limit in Eqs.~\eqref{mean-diff-an-ref}~and~\eqref{var-an-diff-ref}, we obtain 
\begin{align}
    &\mu^{\text{diff}}_{\text{an}} (L,t\ll L^2/D)\approx  \frac{ \rho  \sqrt{D t}}{\sqrt{\pi }}, \label{mean-diff-asym-short-ref} \\
    & 
    {\sigma^{\text{diff}}_{\text{an}}(L,t\ll L^2/D)}^2\approx  \frac{ \rho  \sqrt{D t}}{\sqrt{\pi }}. \label{var-diff-asym-short-ref}
\end{align} 
At very short time scales, the mean and the variance in the annealed setting are the same and also correspond to the infinite system results. However, at larger time scales $t\gg L^2/D$, the mean saturates to the value $N$ and the variance goes to zero as
\begin{align}
   & \mu^{\text{diff}}_{\text{an}} (L,t\gg L^2/D)\approx N -\frac{\rho L^2}{ \sqrt{\pi }   \sqrt{D t}}, \label{mean-diff-asym-large-ref} \\
   & {\sigma^{\text{diff}}_{\text{an}}(L,t\gg L^2/D)}^2 \approx   \frac{\rho L^2}{\sqrt{\pi }   \sqrt{D t}}.\label{var-diff-asym-large-ref}
\end{align}
\begin{figure}
    \centering
    \includegraphics[width=8.5cm]{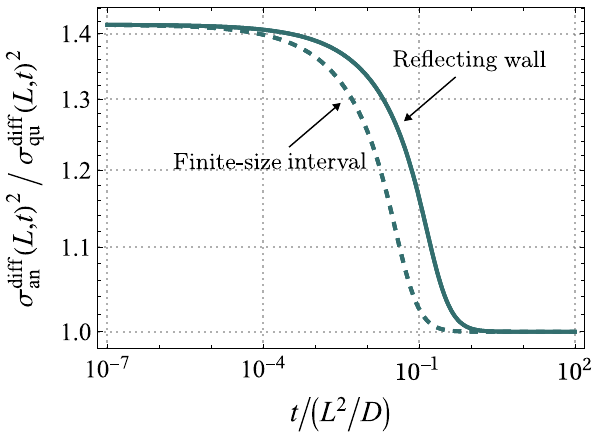}
    \caption{The ratio of the variance of $Q$ in the annealed and quenched settings plotted against rescaled time $t/(L^2/D)$ for two different set-ups:~(a)~in the presence of a reflecting wall (the solid line) and~(b)~ when both sides are open (the dashed line). The variance in the annealed setting is given by \eref{var_two_walls_an} and the variance in the quenched setting has been numerically evaluated using \eref{var_two_walls_qu}, for $L=10,10^2,10^3$ and $D=1$. When $t\ll L^2/D$, the ratio is $\sqrt{2}$, and when $t\gg L^2/D,$ the variances in the annealed and quenched settings become exactly equal and the ratio becomes unity.}
    \label{fig:ratio_diff}
\end{figure}

In deriving the asymptotic time limits we have used the following properties of the complementary error function
\begin{align}
     \begin{array}{l}
       \text{erfc}(z) \approx \left\{\begin{array}{lll}
     1-\frac{2 z}{\sqrt{\pi }}, & \text{when }  z\to 0, \\
      \frac{e^{-z^2}}{\sqrt{\pi } z},  &   \text{when }  z\to \infty.
          \end{array}\right.\end{array} 
\end{align}

At very short times, particles that are close to the boundary at $x=0$ can only get out of the region $[-L,0]$. Meanwhile, particles that are situated near the boundary at $x=-L$ do not get sufficient time to escape through the origin. In effect, the finite size of the system does not come into the picture at very short times. Consequently, the results obtained match with those obtained for the case of an infinite system. Conversely, as time progresses, particles in the bulk or near the reflecting wall at $x=-L$ have sufficient time to exit the box through the origin. It is expected that eventually, all $N$ particles will exit this region, resulting in a mean current of $N$. At these large time scales, with all particles leaving the box, the variance tends to approach zero.

\begin{figure*}
    \centering
    \includegraphics[width=17.5cm]{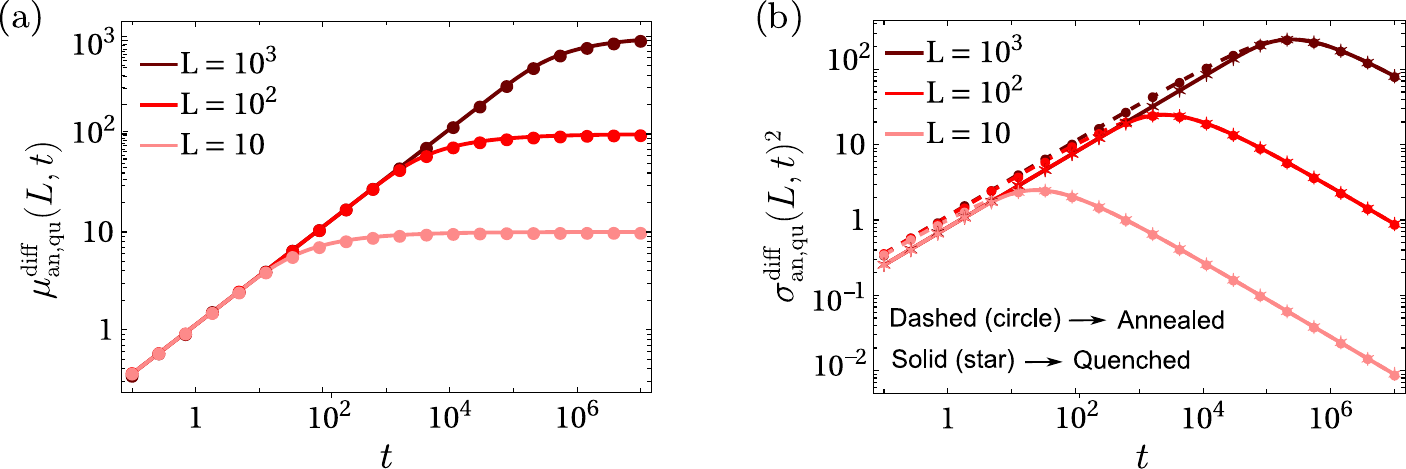}
    \caption{Behavior of the~(a)~mean~and the~(b)~variance of $Q$ as a function of time when both the boundaries at $x=0,-L$ are open for a diffusive system. The mean in the annealed and quenched settings are the same. The variance in the quenched setting (solid curves) differs from the annealed (dashed curves) by a factor of $\sqrt{2}$ at times $t\ll L^2/D$, however, they become equal at times $t\gg L^2/D$. The parameter values used are $\rho=1, D=1$. The stars and circles represent the results obtained through numerical simulations of the microscopic model for the quenched and annealed settings respectively.}
    \label{fig:diff_open}
\end{figure*}


\subsubsection{Quenched setting}
From \eref{mean_two_walls_qu} we see that the mean in the quenched setting is the same as the annealed setting. Therefore we obtain
\begin{align}
\mu^{\text{diff}}_{\text{qu}}(L,t)=\mu^{\text{diff}}_{\text{an}}(L,t),
\end{align}
with the limiting behaviors given in \eref{mean-diff-asym-short-ref} and \eref{mean-diff-asym-large-ref}.

Calculating the variance in the quenched setting is challenging because the integral in \eref{var_two_walls_qu} cannot be explicitly computed. However, we can determine the asymptotic behaviors of the variance of $Q$ using simple arguments. At short times ($t\ll L^2/D$), the system does not experience the effects of finite size and the results obtained are similar to those obtained for infinite systems (as also seen for the annealed case). We thus take the limit $L\to \infty$ in \eref{u-ref-diff} to obtain
\begin{eqnarray}
   U(x_i,t) &\xrightarrow[L \rightarrow \infty]{}&\frac{1}{2} \left(1+\text{erf}\left(\frac{x_i}{2 \sqrt{D t}}\right)\right).
\end{eqnarray} 
Using this result, the integration in \eref{var_two_walls_qu} can be easily performed to obtain
\begin{align}
    {\sigma^{\text{diff}}_{\text{qu}}(L,t\ll L^2/D)}^2 \approx  \frac{ \rho  \sqrt{D t}}{\sqrt{2 \pi }}. \label{var-diff-asym-short-ref-qu}
\end{align}
Similarly, at very large times $t\gg L^2/D$, one can take the limit $L\to 0$ to obtain
\begin{eqnarray}
U(x_i,t)&\xrightarrow[L \rightarrow 0]{}&1-\frac{L e^{-\frac{x_i^2}{4 D t}}}{\sqrt{\pi } \sqrt{D t}}.
\end{eqnarray} 
We next compute the integral in \eref{var_two_walls_qu} in this limit yielding
\begin{align}
     {\sigma^{\text{diff}}_{\text{qu}}(L,t\gg L^2/D)}^2 \approx   \frac{\rho L^2}{ \sqrt{\pi }   \sqrt{D t}}.\label{var-diff-asym-large-ref-qu}
\end{align}
Note that, at time scales where finite size effects are not present ($t\ll L^2/D$), the variance for the quenched setting given in \eref{var-diff-asym-short-ref-qu} is suppressed by a factor of $\sqrt{2}$ compared to the annealed setting provided in \eref{var-diff-asym-short-ref}. However at time scales $t\gg L^2/D$, the finite size effects are dominant and the variance in the quenched and annealed settings become exactly equal to each other. \fref{fig:diff_ref} shows the behavior of the mean and the variance as a function of time for both annealed and quenched settings. The mean is given by Eq.~\eqref{mean-diff-an-ref} for both annealed and quenched settings. The variance is given by~\eref{var_two_walls_an} in the annealed setting and by \eref{var_two_walls_qu} in the quenched setting. Fig.~\ref{fig:ratio_diff} displays the plot of the ratio of the variance in the annealed and quenched settings as a function of the rescaled time $t/(L^2/D)$ for different system sizes $L=10,10^2$ and $10^3$. All the curves for different system sizes collapse into a single curve.  At time scales $t\ll L^2/D$, finite size effects can be neglected and the ratio is close to $\sqrt{2}$. However at large time scales $t\gg L^2/D$, the finite size effects become prominent. Consequently, the annealed and the quenched averages become the same, and the ratio becomes one.

\subsection{Finite size interval}
We next focus on the case where there is no reflecting wall in the system so that particles can escape through either of the boundaries at $x=0$ or $x=-L$. The diffusion propagator in this case is given by
\begin{align}
  G(x,t|x_i)=  \frac{1}{\sqrt{4 \pi D t}}e^{-\frac{(x-x_i)^2}{4D t}}.
\end{align}
Substituting this expression in \eref{u}, we obtain
\begin{align}
    U(x_i,t)=\frac{1}{2} \left(1+\text{erf}\left(\frac{x_i}{2 \sqrt{D t}}\right)+\text{erfc}\left(\frac{L+x_i}{2 \sqrt{D t}}\right)\right). \label{udiff}
\end{align}
We next focus on annealed and quenched settings separately.

\subsubsection{Annealed setting}
We substitute \eref{udiff} in the expression for mean provided in \eref{mean_two_walls}. This yields the exact expression for the mean in the annealed setting as
\begin{align}
    &\mu^{\text{diff}}_{\text{an}}(L,t)\nonumber\\
    &=\underbrace{\frac{2\rho\sqrt{Dt}}{\sqrt{\pi}}}_{\text{infinite size limit}}+ \underbrace{\rho L\left(\text{erfc}\left(\frac{L}{2 \sqrt{D t}}\right)-\frac{2\sqrt{Dt}}{\sqrt{\pi}L}e^{-\frac{L^2}{4Dt}}\right)}_{\text{finite size correction}} .\label{mean-diff-an}
\end{align}
In the asymptotic limit, we obtain the simplified expressions
\begin{align}
     \begin{array}{l}
         \mu^{\text{diff}}_{\text{an}}(L,t)\approx \left\{\begin{array}{lll}
      \frac{2 \rho  \sqrt{D t}}{\sqrt{\pi }}, &   t\ll L^2/D, \\
        N -\frac{\rho L^2}{2 \sqrt{\pi }   \sqrt{D t}},  &   t\gg L^2/D.
          \end{array}\right.\end{array} \label{diff-mean-asy-ref}
\end{align}

The expression for the variance of $Q$ can now be exactly computed 
using \eref{mean-diff-an} and \eref{var-an-diff-ref}. Since this expression is quite long, we do not quote it here. \fref{fig:diff_open} shows the behavior of the mean and the variance as a function of time obtained from Eqs.~\eqref{mean-diff-an}~and~\eqref{var-an-diff-ref} for different system sizes keeping the density $\rho=1$ fixed. In the asymptotic limits, we obtain the simple expressions,

\begin{align}
     \begin{array}{l}
       \sigma^{\text{diff}}_{\text{an}}(L,t)^2 \approx \left\{\begin{array}{lll}
      \frac{2 \rho  \sqrt{D t}}{\sqrt{\pi }}, &   t\ll L^2/D, \\
       \frac{\rho L^2}{2 \sqrt{\pi }   \sqrt{D t}},  &   t\gg L^2/D.
          \end{array}\right.\end{array}  \label{diff-var-asy-ref}
\end{align}
At short times, the variance is larger by a factor of $2$ as compared to the case with a single reflecting boundary in the annealed setting. 
However, at large times, the variance is lesser by a factor of $2$ as compared to the previous case. 

\subsubsection{Quenched setting}
The expression for the mean in the quenched setting is the same as the annealed setting and is given in \eref{mean-diff-an}. It is difficult to compute the exact closed-form expression for the variance in the quenched setting using the expression for $U(x_i,t)$ provided in \eref{udiff}.
Nevertheless, it is possible to perform a careful asymptotic analysis in the Laplace space (details given \aref{appendix_diffusion}) which yields
\begin{align}
     \begin{array}{l}
          \sigma^{\text{diff}}_{\text{qu}}(L,t)^2\approx \left\{\begin{array}{ll}
     \frac{\sqrt{2} \rho  \sqrt{D t}}{\sqrt{\pi }}, &   t\ll \frac{L^2}{D}, \\
      \frac{\rho L^2}{2 \sqrt{\pi }  \sqrt{D t}},  & t\gg \frac{L^2}{D} .
          \end{array}\right.\end{array}
\end{align}
Similar to the case with a reflecting wall, the variance in the annealed and quenched settings are distinct at short times ($t\ll L^2/D$) and become exactly equal to each other at times $t\gg L^2/D$. \fref{fig:diff_open} shows the behavior of the mean and the variance as a function of time for both annealed and quenched settings. The mean is given by Eq.~\eqref{mean-diff-an} for both annealed and quenched settings. The variance is given by~\eref{var_two_walls_an} in the annealed setting and by \eref{var_two_walls_qu} in the quenched setting. In \fref{fig:ratio_diff}, we display a plot of the exact ratio of the annealed to quenched variance. As before, even in the quenched setting, the variance of $Q$ is larger by a factor of $2$ compared to the case with a single reflecting boundary at short times. However, at large times, the variance is lower by a factor of $2$. This demonstrates how boundary conditions can influence the transport properties of stochastic systems over time.

\section{Run-and-tumble particles}
\label{sec_rtp}


\begin{table}[t!]
\centering
{\setlength{\extrarowheight}{7pt}%
\begin{tabular}{c|ccc}
& $t \rightarrow 0$ & $t \rightarrow \infty$            \\ [7pt]
\hline
 \multirow{2}{*}{RTP reflecting boundary} & $ \frac{1}{2}\rho vt$   & $   \frac{\rho L^2}{ \sqrt{\pi }   \sqrt{D_{\text{eff}} t}}$   & $\sigma^{\text{rtp}}_{\text{an}}(L,t)^2$  \\ [7pt]
 \cline{2-4}
 & $\frac{1}{4}\rho vt$   & $\ \frac{\rho L^2}{ \sqrt{\pi }   \sqrt{D_{\text{eff}} t}}$   & $\sigma^{\text{rtp}}_{\text{qu}}(L,t)^2$  \\  [7pt]
   \hline
   \multirow{2}{*}{RTP finite interval} & $\rho vt$   & $\frac{1}{2}  \frac{\rho L^2}{ \sqrt{\pi }   \sqrt{D_{\text{eff}} t}}$   & $\sigma^{\text{rtp}}_{\text{an}}(L,t)^2$ \\ [7pt]
       \cline{2-4}
 & $\frac{1}{2}\rho vt$  & $\frac{1}{2}\frac{\rho L^2}{ \sqrt{\pi }   \sqrt{D_{\text{eff}} t}}$ & $\sigma^{\text{rtp}}_{\text{qu}}(L,t)^2$   \\ 
\end{tabular}}
 \caption{Asymptotic behavior of current fluctuations for run and tumble particle motion in the annealed and quenched settings.}
\label{table_RTP}
\end{table}
\begin{figure*}
    \centering
    \includegraphics[width=17.5cm]{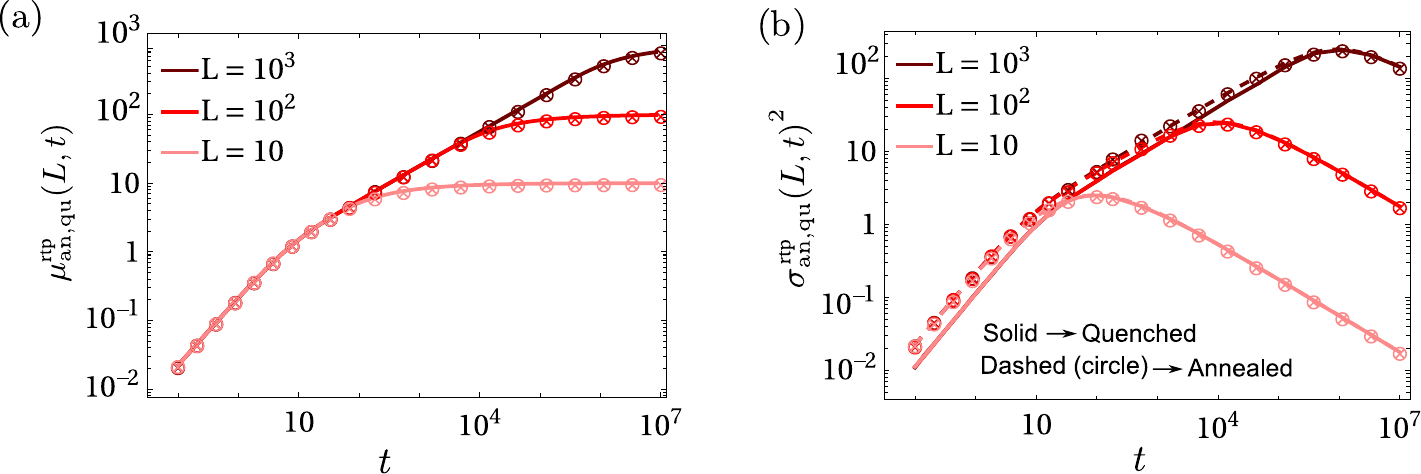}
    \caption{Behavior of the~(a)~mean~and the~(b)~variance in both annealed and quenched settings for RTP in the presence of a reflecting wall at $x=-L$. The mean in the annealed and quenched settings are the same. The variance in the quenched setting (solid curves) differs from the annealed (dashed curves) by a factor of ${2}$ at times $t\ll L^2/D_{\text{eff}}$, however, they become equal at times $t\gg L^2/D_{\text{eff}}$. The parameter values used are $v=\sqrt{0.2},\gamma=0.1$ to have $D_{\text{eff}}=v^2/2\gamma=1$.  The circles with crosses represent the results obtained through numerical simulations of the microscopic model and the dashed curves represent the results obtained through numerical Laplace inversion of \eref{rtp-mean-an-ref} along with \eref{var_two_walls_an} for the annealed setting. The variance in quenched setting (solid curves) has been obtained through microscopic simulations.}
    \label{fig:rtp_ref}
\end{figure*}

In this section, we focus on a system of non-interacting run-and-tumble particles in one dimension. We analyze the statistics of the integrated current $Q$ in a one-dimensional system of non-interacting run-and-tumble particles (RTPs). The dynamics of an RTP consist of run and tumble phases. During the run phase, the particle moves with a constant velocity $v$ and during the tumble phase, the particle instantaneously changes its direction of velocity. The Langevin equation   governing the motion of an RTP can be written as
\begin{align}
    \frac{dx}{dt}=v \sigma(t),
\end{align}
where $\sigma(t)=\pm 1$ is a dichotomous noise and it switches between the two values after a random time $\tau$ which is distributed according to an exponential distribution $p(\tau)=\gamma e^{-\gamma \tau}$. The asymptotic behaviors of current fluctuations for non-interacting RTPs in various settings are summarized in Table~\ref{table_RTP}.

As for the diffusive case, we first focus on the situation with a reflecting wall at $x=-L$.
\subsection{One reflecting wall}
The boundary conditions for an RTP in the presence of a reflecting wall have to be defined carefully. After a reflection from the wall, each particle has two possibilities for its orientation, (i) it continues to move in the same direction i.e. towards the wall or (ii) it changes the orientation after reflection and starts moving away from the wall. 
In this paper, we consider the latter case where velocity is reversed after each reflection. This prevents the accumulation of particles near the wall~\cite{angelani2023one}. The Green's function for RTP has a simple form in the Laplace space. The Laplace transform of a function $f(t)$ is defined as $\widetilde f(s)=\int_0^t dt f(t) e^{-st}$. The propagator of RTP can be computed using the image method 
as~\cite{feller1991introduction}
\begin{align}
  &\widetilde{G}(x,s|x_i)\nonumber\\
  &= \frac{\sqrt{s(s+2\gamma)}}{2vs}\left(e^{-\frac{\sqrt{s(s+2\gamma)}}{v}|x-x_i|}+ e^{-\frac{\sqrt{s(s+2\gamma)}}{v}|x+2L+x_i|}\right). 
  \end{align}
Using \eref{u-ref}, we next compute the Laplace transform of $U(x_i,t)$ as,
  \begin{align}
      \widetilde{U}(x_i,s)=\frac{e^{\frac{ x_i \sqrt{s (2 \gamma +s)}}{v}}}{2s}+\frac{e^{-\frac{\left(2 L+x_i\right) \sqrt{s (2 \gamma +s)}}{v}}}{2s}. \label{u-rtp-ref}
  \end{align}
In the subsequent sections, we focus on the annealed and quenched settings separately.
\subsubsection{Annealed setting}

We take a Laplace transform of the expression for the mean provided in Eq.~\eqref{mean_two_walls} to obtain 
\begin{align}
    \widetilde  \mu_{\text{an}}(L,s)=\rho \int_{-L}^0 \widetilde U(z,s)dz.
    \label{mean_two_walls_ls}
\end{align}
Substituting \eref{u-rtp-ref} in the above equation, we obtain the expression for the mean of $Q$ in Laplace space as
\begin{align}
{\widetilde\mu}^{\text{rtp}}_{\text{an}}(L,s)=\underbrace{\frac{\rho v}{2 s \sqrt{s (2 \gamma +s)}}}_{\text{infinite size limit}} - \underbrace{\frac{\rho v e^{-\frac{2 L \sqrt{s (2 \gamma +s)}}{v}}}{2 s \sqrt{s (2 \gamma +s)}}}_{\text{finite size correction}}. \label{rtp-mean-an-ref}
\end{align}
The first term in the above expression represents the infinite size ($L\to \infty$) limit while the second term is a finite size correction. Since the exact inversion of the above expression is difficult, we focus on the asymptotic behaviors taking different limits of $s$ as explained below.

For RTPs, there are two important time scales,~(i)~one timescale is associated with the mean run time $t=1/\gamma$ between consecutive tumbles and~(ii)~the other timescale is associated with the finite size of the system $t=L^2/D_{\text{eff}}$ where $D_{\text{eff}}=v^2/2\gamma$ is the effective diffusion constant for an RTP in one dimension. At very large times, $t\gg 1/\gamma$ the statistical properties of an RTP become similar to that of a Brownian particle with an effective diffusion constant $D_{\text{eff}}$. In this paper, we consider the case where $L^2/D_{\text{eff}} \gg 1/\gamma$. Thus the limit $s\to\infty$ corresponds to timescales $t\ll L^2/D_{\text{eff}}$. In this limit, we observe that the second term in \eref{rtp-mean-an-ref} is exponentially suppressed as compared to the first term and we obtain
\begin{eqnarray}
     {\widetilde\mu}^{\text{rtp}}_{\text{an}}(L,s)&\xrightarrow[s\rightarrow \infty]{}&\frac{\rho v  }{2s \sqrt{s (2 \gamma +s)}},
    \end{eqnarray}
  which upon inversion yields 
  \begin{equation}
{\mu}^{\text{rtp}}_{\text{an}}(L,t\ll L^2/D_{\text{eff}} ) =\frac{\rho v t}{2}  e^{-\gamma t } (I_0(t \gamma )+I_1(t \gamma )),
\label{meant-rtp-ref}
 \end{equation}
where $I_0(z)$ and $I_1(z)$ are the modified Bessel functions of the first
kind. The asymptotic behaviors of the modified Bessel function of the first kind (and order $\nu$) are given as
\begin{align}
     \begin{array}{l}
       I_{\nu}(z) \approx \left\{\begin{array}{lll}
    z^{\nu } \left(\frac{2^{-\nu }}{\Gamma (\nu +1)}+\frac{2^{-\nu -2} z^2}{(\nu +1) \Gamma (\nu +1)}\right), & \text{when }  z\to 0, \\
     \frac{e^z}{\sqrt{2 \pi } \sqrt{z}},  &   \text{when }  z\to \infty.
          \end{array}\right.\end{array} 
\end{align}
Substituting these expressions in \eref{meant-rtp-ref}, one obtains the limiting behaviors of the mean of $Q$ as 
\begin{align}
     \begin{array}{l}
        {\mu}^{\text{rtp}}_{\text{an}}(L,t)\approx \left\{\begin{array}{lll}
       \frac{1}{2}\rho vt, &   t\ll 1/\gamma, \\
         \frac{\rho\sqrt{D_{\text{eff}}t}}{\sqrt{\pi}},  &   \frac{1}{\gamma}\ll t \ll \frac{L^2}{D_{\text{eff}}}.
          \end{array}\right.\end{array}
\end{align}
 To obtain the large time ($t\gg L^2/D_{\text{eff}}$) behavior, we take the $s\to 0$ limit of the expression provided in \eref{rtp-mean-an-ref} yielding
\begin{eqnarray}
      {\widetilde\mu}^{\text{rtp}}_{\text{an}}(L,s)&\xrightarrow[s\rightarrow 0]{}&\frac{N }{s}-\frac{ \rho L^2  }{\sqrt{D_{\text{eff}}} \sqrt{s} },
\end{eqnarray}
  which upon inversion yields 
  \begin{equation}
{\mu}^{\text{rtp}}_{\text{an}}(L,t\gg L^2/D_{\text{eff}} ) =N-\frac{ \rho L^2  }{\sqrt{\pi } \sqrt{D_{\text{eff}} t} }.
\label{mean_rtp_ref_lt}
\end{equation}

The asymptotic behavior of the variance can be found by substituting the asymptotic expressions for the mean provided in Eqs.~\eqref{meant-rtp-ref}~and~\eqref{mean_rtp_ref_lt} directly in \eref{var_two_walls_an}. This yields
\begin{align}
     \begin{array}{l}
         \sigma^{\text{rtp}}_{\text{an}}(L,t)^2\approx \left\{\begin{array}{lll}
       \frac{1}{2}\rho vt, &   t\ll 1/\gamma, \\
         \frac{\rho \sqrt{D_{\text{eff}}t}}{\sqrt{\pi}},  &   \frac{1}{\gamma}\ll t \ll \frac{L^2}{D_{\text{eff}}},\\
         \frac{\rho L^2}{ \sqrt{\pi }   \sqrt{D_{\text{eff}} t}} & t\gg L^2/D_{\text{eff}} .
          \end{array}\right.\end{array}
          \label{var_mean_ref_an_alt}
\end{align}
At time scales $t\gg 1/\gamma$, the mean and the variance behave similar to that of the diffusive case as in \eref{mean-diff-asym-short-ref}-(\ref{var-diff-asym-large-ref}) with the diffusion constant $D$ replaced by $D_{\text{eff}}$. \fref{fig:rtp_ref} shows the behavior of the mean and the variance as a function of time obtained through numerical inversion of Eq.~\eqref{rtp-mean-an-ref}~and~using these results in \eref{var_two_walls_an}.
\begin{figure}
    \centering
    \includegraphics[width=8.5cm]{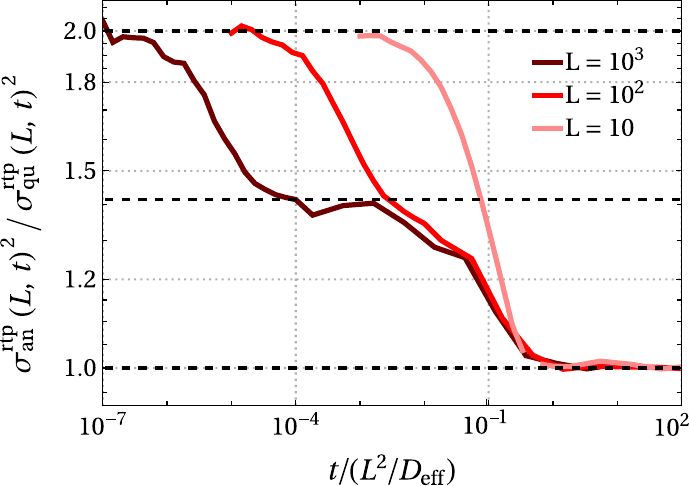}
    \caption{Ratio of the variances of $Q$ in the annealed and quenched settings for RTPs in the presence of a reflecting wall. For small timescales ($t\ll 1/\gamma$), the ratio is $2$. With time it starts decreasing and at intermediate time scales $1/\gamma \ll t \ll L^2/D_{\text{eff}}$, the ratio saturates to the value $\sqrt{2}$. This saturation is more evident for system size $L=10^3$ as the intermediate region is broad here. At time scale $t\gg L^2/D_{\text{eff}}$ all the curves merge and eventually saturate to unity. While computing the ratio, the numerator has been obtained through numerical Laplace inversion of \eref{rtp-mean-an-ref} along with \eref{var_two_walls_an} while the denominator has been estimated using microscopic simulations.}
    \label{fig:ratio_rtp_ref}
\end{figure}

\subsubsection{Quenched setting}
Similar to the case of diffusion, the mean in the quenched setting is the same as that in the annealed setting. That is,
\begin{align}
\mu^{\text{rtp}}_{\text{qu}}(L,t)=\mu^{\text{rtp}}_{\text{an}}(L,t).
\label{equality_mean_rtp}
\end{align}
The exact asymptotic behaviors of the mean are provided in Eqs.~\eqref{meant-rtp-ref}~and~\eqref{mean_rtp_ref_lt}. The asymptotic limits for the variance can be computed using similar arguments we applied for the diffusive case. At very short times $t\ll L^2/D_{\text{eff}}$, we take the limit $L\to \infty$ in \eref{u-rtp-ref} and follow a similar calculation as we did for the diffusive case to obtain
\begin{align}
     & \sigma^{\text{rtp}}_{\text{qu}}(L,t\ll L^2/D_{\text{eff}})^2=\frac{\rho v}{8}te^{-2\gamma t}\big[(4+\pi L_0(2\gamma t))I_1(2\gamma t) \nonumber\\
     &\hspace{3cm}+ (2-\pi L_1(2\gamma t))I_0(2\gamma t) \big],
\end{align}
where $L_0(z), L_1(z)$ are the modified Struve functions. 
Further depending on the time scale $ 1/\gamma$, we obtain the limiting behaviors for the variance of $Q$
\begin{align}
     \begin{array}{l}
        \sigma^{\text{rtp}}_{\text{qu}}(L,t)^2 \approx \left\{\begin{array}{lll}
       \frac{1}{4}\rho vt, &   t\ll 1/\gamma, \\
         \frac{\rho\sqrt{D_{\text{eff}}t}}{\sqrt{2\pi}},  &   \frac{1}{\gamma}\ll t \ll \frac{L^2}{D_{\text{eff}}}.
          \end{array}\right.\end{array}
\end{align}
Here, we have used the following asymptotic behaviors of the Struve functions
\begin{align}
     \begin{array}{l}
       L_{\nu}(z) \approx \left\{\begin{array}{lll}
   z^{\nu } \left(\frac{2^{-\nu } z}{\sqrt{\pi } \Gamma \left(\nu +\frac{3}{2}\right)}+\frac{2^{-\nu -1} z^3}{3 \sqrt{\pi } \Gamma \left(\nu +\frac{5}{2}\right)}\right), & \text{when }  z\to 0, \\
     \frac{e^z}{\sqrt{2 \pi } \sqrt{z}},  &   \hspace{-0.16cm} \text{when }  z\to \infty.
          \end{array}\right.\end{array} 
\end{align}

The large-time asymptotic behavior of the variance can be computed by taking the limit $L \to \infty$ in \eref{u-rtp-ref} and performing a similar calculation as for the diffusive case, or it can be derived directly from the fact that at this timescale, the statistical properties of an RTP is similar to that of a Brownian particle with a modified diffusion constant $D=D_{\text{eff}}$. We thus obtain
\begin{align}
     \sigma^{\text{rtp}}_{\text{qu}}(t\gg L^2/D_{\text{eff}})^2 =  \frac{\rho L^2}{ \sqrt{\pi }   \sqrt{D_{\text{eff}} t}},
\end{align}
\begin{figure*}
    \centering \includegraphics[width=17.5cm]{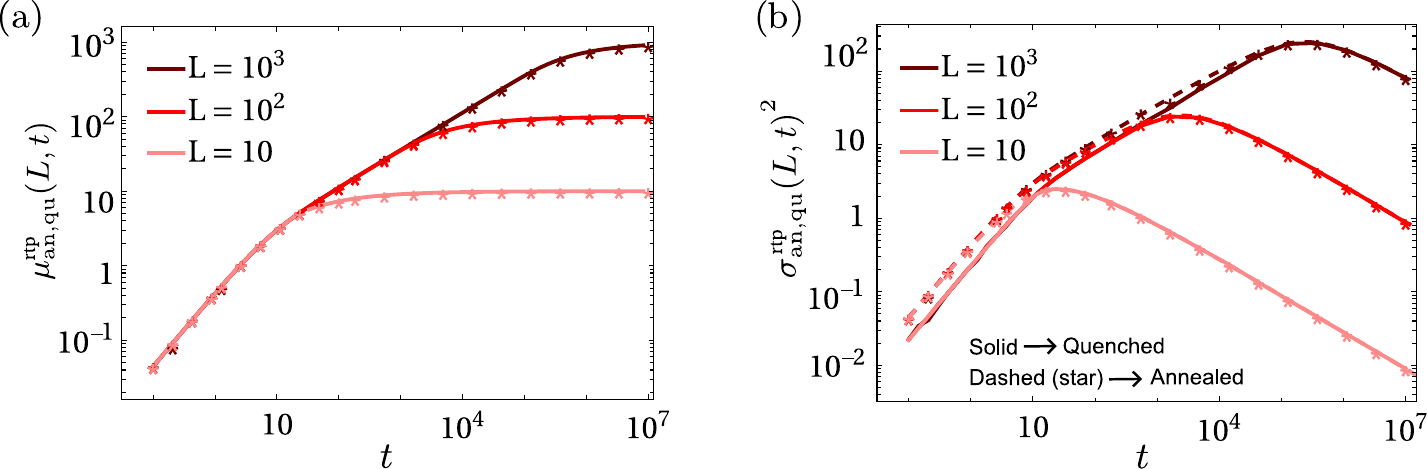}
    \caption{Behavior of the~(a)~mean and the~(b)~variance in both annealed and quenched settings for run-and-tumble particles when both sides at $x=0,-L$ are open. The mean in the annealed and quenched settings are the same. The variance in the quenched setting (solid curves) differs from the annealed (dashed curves) by a factor of ${2}$ at times $t\ll L^2/D_{\text{eff}}$, however, they become equal at times $t\gg L^2/D_{\text{eff}}$. The parameter values used are $v=\sqrt{0.2},\gamma=0.1$ so that $D_{\text{eff}}=v^2/2\gamma=1$. The stars represent the results obtained through numerical simulations of the microscopic model and the dashed curves represent the results obtained through numerical Laplace inversion of \eref{mean-rtp-an-ls} along with \eref{var_two_walls_an} for the annealed setting. The variance in quenched setting (solid curves) has been entirely obtained through microscopic simulations.  
    }
    \label{fig:rtp_open}
\end{figure*}which is the same as the large time behavior of the variance in the annealed setting given in \eref{var_mean_ref_an_alt}.  \fref{fig:rtp_ref} displays the behavior of the mean and the variance as a function of time for both annealed and quenched settings. The mean is given by numerical inversion of Eq.~\eqref{rtp-mean-an-ref} for both annealed and quenched settings. The variance is given by~\eref{var_two_walls_an} in the annealed setting and by \eref{var_two_walls_qu} in the quenched setting. Fig.~\ref{fig:ratio_rtp_ref} displays the plot of the ratio of the variance in the annealed and quenched settings as a function of the rescaled time $t/(L^2/D_{\text{eff}})$ for different system sizes $L=10,10^2$ and $10^3$. Unlike the diffusive case, the curves do not collapse into a single curve as RTPs have different timescales involved in addition to the diffusion timescale. At time scales $t\ll 1/\gamma$, finite size effects can be neglected and the ratio is close to ${2}$. We see that at intermediate time scales $1/\gamma \ll t \ll L^2/D_{\text{eff}}$, the ratio saturates close to the value $\sqrt{2}$. However at large time scales $t\gg L^2/D_{\text{eff}}$, the finite size effects become prominent. Consequently, the annealed and the quenched averages become the same, and the ratio becomes one.


 \subsection{Finite size interval}
 We next focus on the case where the particles can escape either through the boundary at $x=0$ or $x=-L$. The propagator for an RTP in the Laplace space is given as~\cite{banerjee2020current}
 \begin{align}\label{rtp-prp}
    &\widetilde{G}(x,s|x_i)= \frac{\sqrt{s(s+2\gamma)}}{2vs}e^{-\frac{\sqrt{s(s+2\gamma)}}{v}|x-x_i|}.
\end{align}
Substituting this expression in \eref{u}, we obtain 
\begin{align}
    \widetilde{U}(x_i,s)=\frac{e^{\frac{x_i \sqrt{s (2 \gamma +s)}}{v}}}{s}+\frac{e^{\frac{-(L+x_i) \sqrt{s (2 \gamma +s)}}{v}}}{s}.\label{u-rtp}
\end{align}
We next focus on the cases of annealed and quenched averages separately.
\subsubsection{Annealed setting}
We first focus on the annealed setting where the positions of the particles are allowed to fluctuate initially. Substituting \eref{u-rtp} in \eref{mean_two_walls_ls} we obtain the expression for the mean in Laplace space as
\begin{align}
  {\widetilde\mu}^{\text{rtp}}_{\text{an}}(L,s)=\underbrace{\frac{\rho v  }{s \sqrt{s (2 \gamma +s)}}}_{\text{infinite size limit}}-\underbrace{\frac{\rho v  e^{-\frac{ L\sqrt{s (2 \gamma +s)}}{  v}}}{s \sqrt{s (2 \gamma +s)}}}_{\text{finite size correction}}.\label{mean-rtp-an-ls}
\end{align}
Using this expression, the asymptotic behaviors of the mean and the variance in real-time can be computed as before. For the mean, we obtain
\begin{align}
     \begin{array}{l}       \mu^\text{rtp}_{\text{an}}(L,t)\approx \left\{\begin{array}{lll}
       \rho vt, &   t\ll 1/\gamma, \\
         \frac{2\rho \sqrt{D_{\text{eff}}t}}{\sqrt{\pi}},  &   \frac{1}{\gamma}\ll t \ll \frac{L^2}{D_{\text{eff}}},\\
        N-\frac{ \rho L^2  }{2\sqrt{\pi } \sqrt{D_{\text{eff}} t} } & t\gg L^2/D_{\text{eff}} .
          \end{array}\right.\end{array}
\end{align}
and for the variance, we obtain
\begin{align}
     \begin{array}{l}
         \sigma^{\text{rtp}}_{\text{an}}(L,t)^2\approx \left\{\begin{array}{lll}
       \rho vt, &   t\ll 1/\gamma, \\
         \frac{2\rho \sqrt{D_{\text{eff}}t}}{\sqrt{\pi}},  &   \frac{1}{\gamma}\ll t \ll \frac{L^2}{D_{\text{eff}}},\\
         \frac{\rho L^2}{2 \sqrt{\pi }   \sqrt{D_{\text{eff}} t}} & t\gg L^2/D_{\text{eff}} .
          \end{array}\right.\end{array}
          \label{var_an_rtp_finite_limits}
\end{align}


\begin{figure}
    \centering
    \includegraphics[width=8.5cm]{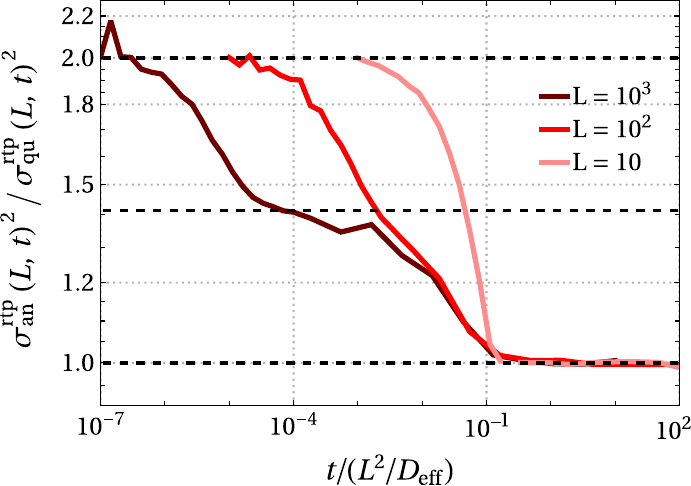}
    \caption{Ratio of the variance of $Q$ in the annealed and quenched settings for RTPs in a finite-sized interval with open boundaries. The variance in the annealed setting has been obtained through numerical Laplace inversion of \eref{mean-rtp-an-ls} along with \eref{var_two_walls_an}. The variance in the quenched setting has been obtained through numerical simulations of the microscopic model. Unlike the Brownian case, the curves for different 
 system sizes $L$ do not collapse into a single curve at short and intermediate times. However, when $t\gg L^2/D_{\text{eff}}$, all the curves merge and saturate to the value $1$.}
    \label{fig:ratio_rtp_open}
\end{figure}

\subsubsection{Quenched setting}
The mean in the quenched setting is the same as the mean in the annealed setting
and is given in \eref{mean-rtp-an-ls}. Since it is difficult to find the exact expression of the variance in the quenched setting in closed form, we focus on the asymptotic behaviors. A careful asymptotic analysis in the Laplace space (details given in \aref{appendix_rtp}) yields

\begin{align}
     \begin{array}{l}
          \sigma^{\text{rtp}}_{\text{qu}}(L,t)^2\approx \left\{\begin{array}{ll}
   \frac{1}{2}{\rho v t}, &   t\ll \frac{1}{\gamma}, \\
      \frac{\rho L^2}{2\sqrt{ \pi D_{\text{eff}}t}},  & t\gg \frac{L^2}{D_{\text{eff}}} .
          \end{array}\right.\end{array}
\end{align}

\fref{fig:rtp_open} displays the behavior of the mean and the variance as a function of time for both annealed and quenched settings. The mean is given by numerical inversion of Eq.~\eqref{mean-rtp-an-ls} for both annealed and quenched settings. The variance is given by~\eref{var_two_walls_an} in the annealed setting and by \eref{var_two_walls_qu} in the quenched setting. In \fref{fig:ratio_rtp_open}, we display a plot of the ratio of the variance in the annealed and quenched settings as a function of the rescaled time $t/(L^2/D_{\text{eff}})$ for different system sizes $L=10,10^2$ and $10^3$. Similar to the set-up with a reflecting wall, the curves do not collapse into a single curve as RTPs have different timescales involved in addition to the diffusion timescale. At time scales $t\ll 1/\gamma$, the ratio is close to ${2}$. At intermediate time scales $1/\gamma \ll t \ll L^2/D_{\text{eff}}$, the ratio is close to the value $\sqrt{2}$. However at large time scales $t\gg L^2/D_{\text{eff}}$, the finite size effects become prominent and the ratio saturates to $1$. Compared to the reflecting case, the variance is larger by a factor of $2$ at times $t\ll L^2/D_{\text{eff}}$. However, at large times, the variance gets suppressed by a factor of $2$ as compared to the reflecting case. This is exactly the same behavior we observed for the system of diffusing particles. This demonstrates how boundary conditions can influence the transport properties of stochastic systems over time. A more intricate understanding of these various factors would require a detailed study of current fluctuations in different system geometries across various spatial dimensions.


\section{Discussion and outlook}
\label{sec_discussion}

In this paper, we have studied the fluctuations in the number of particles exiting the boundaries of a finite-sized one-dimensional box. We investigated specific examples of passive as well as active systems; namely non-interacting diffusive and run-and-tumble particles respectively. We demonstrated how various initial conditions, system geometry, and boundary conditions affect the transport properties of these systems over time. 

For the system of diffusive particles, we showed that the ratio of fluctuations in the annealed and quenched settings changes from a value of $\sqrt{2}$ at short times ($t \ll {L^2}/{D}$) to $1$ at large times ($t \gg {L^2}/{D}$). While for run-and-tumble particles, this ratio changes from a value of $2$ at short times ($t \ll 1/\gamma$) to $1$ at large times ($t \gg {L^2}/{D_{\text{eff}}}$) through an intermediate saturation regime where the ratio takes up the value $\sqrt{2}$. This intermediate saturation regime corresponds to the time scale ${1}/{\gamma}\ll t \ll {L^2}/{D_{\text{eff}}}$ at which the dynamics of run-and-tumble particles becomes effectively diffusive. The timescale at which the ratio saturates to $1$ is the diffusive timescale which goes as $t \approx L^2/D$ for diffusive systems and $t \approx L^2/D_{\text{eff}}$ for active systems, where $D_{\text{eff}}$ is the effective diffusion constant for run-and-tumble particles in one dimension. 

Interestingly, we demonstrated that the boundary conditions also play a crucial role in determining the dynamic behavior of current fluctuations. The setup with two open boundaries displays larger fluctuations by a factor of $2$ at short times compared to the setup with only one open boundary. However, the former setup exhibits lesser fluctuations by the same factor of $2$ at large times. This can be qualitatively understood as follows: At short times, the particles in the setup with two open boundaries have two escape routes, thereby increasing the fluctuations by a factor of $2$. However, at large times, the fluctuations are predominantly determined by single-particle events, and the probability that an unbiased single particle escapes through one of the boundaries is $1/2$. Consequently, this reduces the fluctuations by a factor of $2$. 

Our exact analytical results reveal how slight variations in the initial conditions and system geometry can affect the dynamic behavior of current fluctuations in stochastic systems. 
Our study is a first step towards understanding the effusion of particles through finite-sized regions across different spatial dimensions, which can be investigated using similar methods discussed in this paper. The study of particle effusion has applications in designing membranes and porous materials, where controlled diffusion or leakage plays a pivotal role, as well as in the transportation of ions or molecules across cellular membranes~\cite{karger2008single,post1991diffusion,marbach2021intrinsic}. Naturally, a careful analytical analysis of the problem of effusion through different confining volumes in higher dimensions will help to understand the underlying factors governing current fluctuations. 

It would be intriguing to investigate whether a universal behavior of current fluctuations exists, one that depends on the system's geometry, determined by factors such as the number of reflecting boundaries and available escape routes. Testing the results of this paper using coarse-grained field theories such as macroscopic fluctuation theory (MFT)~\cite{bertini2005current,bertini2006non,bertini2007stochastic,bertini2009towards,derrida2007non,agranov2023macroscopic,jose2023current} is also a worthwhile future investigation. Finally, it would also be interesting to extend the computations presented in this paper to interacting systems such as the symmetric simple exclusion process (SSEP)~\cite{levitt1973dynamics,arratia1983motion,derrida2004current,krapivsky2015tagged,dandekar2022macroscopic} and the ABC model~\cite{clincy2003phase,bodineau2011phase}.


\section{Acknowledgement} 
The numerical calculations reported in
this work were carried out on the Nandadevi cluster, which is maintained and supported by The Institute of Mathematical Science’s High-Performance Computing Center. KR acknowledges funding through intramural funds at TIFR Hyderabad and the SERB-MATRICS grant MTR/2022/000966. AP gratefully acknowledges research support from the Department of Science and Technology, India, SERB Start-up Research Grant Number SRG/2022/000080, and the Department of Atomic Energy, India.


\section*{Appendices}
\appendix

\section{ Current fluctuations for Brownian particles confined in a finite interval - quenched setting }

\label{appendix_diffusion}
For a Brownian particle confined in a finite interval, the function $U(x_i,t)$ can be computed as in Eq.~\eqref{udiff}. In Laplace space, this expression becomes
\begin{equation}
    \widetilde{U}(x_i,s)=\frac{e^{-\sqrt{\frac{s}{D}} \left(L+x_i\right)} \left(1+e^{\sqrt{\frac{s}{D}} \left(L+2
   x_i\right)}\right)}{2 s}.
\end{equation}
The variance in the quenched setting can be computed by taking a Laplace transform of the expression in \eref{var_two_walls_qu}. This yields
\begin{align}
       \widetilde{\sigma}_{\text{qu}}(L,s)^2= \widetilde\mu_{\text{qu}}(L,s)-\rho \int_{0}^Ldz~\mathcal{L}(s)[ U^2(z,t)].
      \label{var_bm_exp_s}
\end{align}
The expression for $\widetilde\mu_{\text{qu}}^{\text{diff}}(L,s)$ for Brownian motion can be computed as
\begin{equation}
 \widetilde\mu_{\text{qu}}^{\text{diff}}(L,s)=\rho \int_{0}^L \widetilde{U}(z,s)dz= \rho \left(1-e^{-L \sqrt{\frac{s}{D}}}\right)
   \sqrt{\frac{D}{s^3}}.
   \label{mean_bm_s}
\end{equation}
The integral in the second term of Eq.~\eqref{var_bm_exp_s} can be computed using the identity
\begin{align}
     &\int_{0}^L dz~  \mathcal{L}(s)[ U^2(z,t)]\nonumber\\
     &=\frac{1}{2\pi}\int_{-\infty}^{\infty}dk\int_{0}^L dz~ \widetilde{U}(z,s/2-ik)\widetilde{U}(z,s/2+ik).
  \label{integral_bm}
\end{align}
We provide a short derivation of this identity below.
\begin{align}
     &\int_{0}^L dz~  \mathcal{L}(s)[ U^2(z,t)]\nonumber\\
     &=\int_{0}^L dz~  \int_0^{\infty} dt e^{-st/2} U(z,t)\times \nonumber\\
     &\hspace{3.6cm}\int_0^\infty dt' \delta(t-t') e^{-st'/2} U(z,t')\nonumber\\
     &=\int_{0}^L dz\int_0^{\infty} dt e^{-st/2} U(z,t) \times\nonumber\\
     &\hspace{1.4cm}\int_0^\infty dt' \left(\frac{1}{2\pi}\int_{-\infty}^{\infty}dk e^{ik(t-t')}\right) e^{-st'/2} U(z,t')\nonumber\\
     &=\frac{1}{2\pi}\int_{-\infty}^{\infty}dk\int_{0}^L dz\int_0^{\infty} dt e^{-st/2} e^{ikt} U(z,t) \times\nonumber\\
     &\hspace{4cm}\int_0^\infty dt'   e^{-st'/2} e^{-ikt'} U(z,t')\nonumber\\
     &=\frac{1}{2\pi}\int_{-\infty}^{\infty}dk\int_{0}^L dz~ \widetilde{U}(z,s/2-ik)\widetilde{U}(z,s/2+ik).
\end{align}

The integral over $z$ in Eq.~(\ref{integral_bm}) can be done explicitly. Since the resultant expression is quite long, we do not quote it here. However, this expression admits scaling forms in the limits, $s \xrightarrow{} 0$ and $s \xrightarrow{} \infty$. 
Let us denote
 \begin{eqnarray}
\widetilde F(k,s)=\int_{0}^L dz~ \widetilde{U}(z,s/2-ik)\widetilde{U}(z,s/2+ik).
\label{fks}
\end{eqnarray}   

Using the substitution $u=k/s$, we obtain the following scaling forms for the function $\widetilde F(k,s)$,
\begin{eqnarray}
\widetilde F(k,s)&\xrightarrow[s \rightarrow 0]{}& C_1(s) G_1 (u),\\
\widetilde F(k,s)&\xrightarrow[s \rightarrow \infty]{}& C_2(s) G_2 (u),
\end{eqnarray}
where 
\begin{equation}
  C_1(s)=\frac{L}{s^2}-\frac{L^2}{\sqrt{D}s^{3/2}} ,
\end{equation}
and
\begin{equation}
  C_2(s)=\frac{\sqrt{D}}{s^{5/2}} .
\end{equation}
The expression in Eq.~\eqref{integral_bm} can now be written as
\begin{eqnarray}
  \int_{0}^L dz~  \mathcal{L}(s)[ U^2(z,t)]&=&\frac{1}{2\pi}\int_{-\infty}^{\infty}dk~\widetilde F(k,s)\nonumber\\&\xrightarrow[s \rightarrow 0]{}&\frac{s~C_1(s)}{2\pi}\int_{-\infty}^{\infty}du~G_1(u).\nonumber\\
  &\xrightarrow[s \rightarrow \infty]{}&\frac{s~C_2(s)}{2\pi}\int_{-\infty}^{\infty}du~G_2(u).\nonumber\\
\end{eqnarray}
It can be shown that the values of the integrals $\int_{-\infty}^{\infty}du~G_1(u)$ and $\int_{-\infty}^{\infty}du~G_2(u)$ appearing in the above expressions are exactly equal to $2 \pi$ and $ \pi (2-\sqrt{2})$ respectively. Finally, we obtain
\begin{eqnarray}
  \int_{0}^L dz~  \mathcal{L}(s)[ U^2(z,t)]&\xrightarrow[s \rightarrow 0]{}&s~C_1(s)\nonumber\\&=&\frac{L}{s}-\frac{L^2}{\sqrt{sD}},
  \label{integral_smalls_limit_bm}
\end{eqnarray}
and
\begin{eqnarray}
  \int_{0}^L dz~  \mathcal{L}(s)[ U^2(z,t)]&\xrightarrow[s \rightarrow 0]{}&{s}\frac{\sqrt{2}-1}{2}~C_2(s)\nonumber\\&=&\frac{2-\sqrt{2}}{2}\frac{\sqrt{D}}{s^{3/2}}.
  \label{integral_larges_limit_bm}
\end{eqnarray}
Using Eq.~\eqref{mean_bm_s}, it can also be shown that
\begin{eqnarray}
\widetilde\mu_{\text{qu}}^{\text{diff}}(L,s)&\xrightarrow[s \rightarrow 0]{}&\rho\left(\frac{L}{s}-\frac{L^2}{2\sqrt{sD}}\right),
\label{mean_smalls_limit_bm}
\end{eqnarray}
\begin{eqnarray}
\widetilde\mu_{\text{qu}}^{\text{diff}}(L,s)&\xrightarrow[s \rightarrow \infty]{}&\frac{\rho\sqrt{D}}{s^{3/2}}.
\label{mean_larges_limit_bm}
\end{eqnarray}
Combining results from Eq.~\eqref{integral_smalls_limit_bm}-\eqref{mean_smalls_limit_bm} in Eq.~\eqref{var_bm_exp_s}, we obtain
\begin{eqnarray}
      \widetilde{\sigma}_{\text{qu}}^{\text{diff}}(L,s)^2&\xrightarrow[s \rightarrow 0]{}&\rho\left(\frac{L}{s}-\frac{L^2}{2\sqrt{sD}}\right)-\rho\left(\frac{L}{s}-\frac{L^2}{\sqrt{sD}}\right)\nonumber\\&=&\frac{\rho L^2}{2\sqrt{sD}},
\end{eqnarray}
and
\begin{eqnarray}
      \widetilde{\sigma}_{\text{qu}}^{\text{diff}}(L,s)^2&\xrightarrow[s \rightarrow \infty]{}&\rho\frac{\sqrt{D}}{s^{3/2}}-\rho\frac{2-\sqrt{2}}{2}\frac{\sqrt{D}}{s^{3/2}}\nonumber\\&=& \frac{\rho\sqrt{D}}{\sqrt{2}s^{3/2}},
\end{eqnarray}
which on  inversion yield
\begin{eqnarray}
      \sigma^{\text{diff}}_{\text{qu}}(L,t)^2&\xrightarrow[t \rightarrow \infty]{}&\frac{\rho L^2}{2\sqrt{ \pi Dt}},
      \label{var_qu_t_inf_bm}
\end{eqnarray}
and

\begin{eqnarray}
      \sigma^{\text{diff}}_{\text{qu}}(L,t)^2&\xrightarrow[t \rightarrow 0]{}&\sqrt{2}\frac{ \rho  \sqrt{D t}}{\sqrt{\pi }}.
      \label{var_qu_t_0_bm}
\end{eqnarray}
The expression in Eq.~\eqref{var_qu_t_inf_bm} is exactly equal to the large time asymptotic expression for the variance in the annealed setting we obtained previously in Eq.~\eqref{diff-var-asy-ref}.



\section{ Current fluctuations for run and tumble particles confined in a finite interval - quenched setting }
\label{appendix_rtp}
For a run-and-tumble particle confined in a finite interval, the function $\widetilde U(x_i,s)$ can be computed as in Eq.~\eqref{u-rtp}. Similar to the case for Brownian motion, the variance in the quenched setting can be computed using the expression in \eref{var_bm_exp_s}. The exact expression for the mean in Laplace space $\widetilde\mu_{\text{qu}}^{\text{rtp}}(L,s)$, is given in Eq.~\eqref{mean-rtp-an-ls}. As for the Brownian case, the integral in the second term in \eref{var_bm_exp_s} can be computed using the identity given in \eref{integral_bm}. After performing the integral over $z$, we obtain scaling forms of the resultant expression in the asymptotic limits, $s \xrightarrow{} 0$ and $s \xrightarrow{} \infty$. Using the substitution $u=k/s$, we obtain the following scaling forms for the function $\widetilde F(k,s)$ defined in \eref{fks},
\begin{eqnarray}
\widetilde F(k,s)&\xrightarrow[s \rightarrow 0]{}& C_1(s) G_1 (u),\\
\widetilde F(k,s)&\xrightarrow[s \rightarrow \infty]{}& C_2(s) G_2 (u),
\end{eqnarray}
where 
\begin{equation}
  C_1(s)=\frac{L}{s^2}-\frac{L^2}{\sqrt{D_{\text{eff}}}s^{3/2}} ,
\end{equation}
and
\begin{equation}
  C_2(s)=\frac{v}{s^3} .
\end{equation}
Thus we obtain
\begin{eqnarray}
  \int_{0}^L dz~  \mathcal{L}(s)[ U^2(z,t)]&=&\frac{1}{2\pi}\int_{-\infty}^{\infty}dk~\widetilde F(k,s)\nonumber\\&\xrightarrow[s \rightarrow 0]{}&\frac{s~C_1(s)}{2\pi}\int_{-\infty}^{\infty}du~G_1(u).\nonumber\\
  &\xrightarrow[s \rightarrow \infty]{}&\frac{s~C_2(s)}{2\pi}\int_{-\infty}^{\infty}du~G_2(u).\nonumber\\
\end{eqnarray}
It can be shown that the values of the integrals $\int_{-\infty}^{\infty}du~G_1(u)$ and $\int_{-\infty}^{\infty}du~G_2(u)$ appearing in the above expressions are exactly equal to $2 \pi$ and $ \pi$ respectively. Finally, we obtain
\begin{eqnarray}
  \int_{0}^L dz~  \mathcal{L}(s)[ U^2(z,t)]&\xrightarrow[s \rightarrow 0]{}&s~C_1(s)\nonumber\\&=&\frac{L}{s}-\frac{L^2}{\sqrt{sD_{\text{eff}}}},
  \label{integral_smalls_limit}
\end{eqnarray}
and
\begin{eqnarray}
  \int_{0}^L dz~  \mathcal{L}(s)[ U^2(z,t)]&\xrightarrow[s \rightarrow \infty]{}&\frac{s}{4}~C_2(s)\nonumber\\&=&\frac{v}{2s^2}.
  \label{integral_larges_limit}
\end{eqnarray}

Using Eq.~\eqref{mean-rtp-an-ls}, it can also be shown that
\begin{eqnarray}
\widetilde\mu_{\text{qu}}^{\text{rtp}}(L,s)&\xrightarrow[s \rightarrow 0]{}&\rho\left(\frac{L}{s}-\frac{L^2}{2\sqrt{sD_{\text{eff}}}}\right),
\label{mean_smalls_limit}
\end{eqnarray}
\begin{eqnarray}
\widetilde\mu_{\text{qu}}^{\text{rtp}}(L,s)&\xrightarrow[s \rightarrow \infty]{}&\frac{\rho v}{s^2}.
\label{mean_larges_limit}
\end{eqnarray}
Combining results from Eq.~\eqref{integral_smalls_limit}-\eqref{mean_smalls_limit} in Eq.~\eqref{var_bm_exp_s}, we obtain
\begin{eqnarray}
     \widetilde {\sigma}_{\text{qu}}^{\text{rtp}}(L,s)^2&\xrightarrow[s \rightarrow 0]{}&\rho\left(\frac{L}{s}-\frac{L^2}{2\sqrt{sD_{\text{eff}}}}\right)-\rho\left(\frac{L}{s}-\frac{L^2}{\sqrt{sD_{\text{eff}}}}\right)\nonumber\\&=&\frac{\rho L^2}{2\sqrt{sD_{\text{eff}}}},
\end{eqnarray}
and
\begin{eqnarray}
     \widetilde {\sigma}_{\text{qu}}^{\text{rtp}}(L,s)^2&\xrightarrow[s \rightarrow \infty]{}&\rho\frac{v}{s^2}-\rho\frac{v}{2s^2}\nonumber\\&=&\frac{\rho v}{2s^2},
\end{eqnarray}
which on  inversion yield
\begin{eqnarray}
      \sigma_{\text{qu}}^{\text{rtp}}(L,t)^2&\xrightarrow[t \rightarrow \infty]{}&\frac{\rho L^2}{2\sqrt{ \pi D_{\text{eff}}t}},
      \label{var_qu_t_inf}
\end{eqnarray}
and

\begin{eqnarray}
      \sigma^{\text{rtp}}_{\text{qu}}(L,t)^2&\xrightarrow[t \rightarrow 0]{}&\frac{\rho v t}{2}.
      \label{var_qu_t_0}
\end{eqnarray}
The expression in Eq.~\eqref{var_qu_t_inf} is exactly equal to the large time asymptotic expression for the variance in the annealed setting we obtained previously in Eq.~\eqref{var_an_rtp_finite_limits}.

\newpage
\bibliography{references}

\end{document}